# Quantal Consequences of Perturbations Which Destroy Structurally Unstable Orbits in Chaotic Billiards

Harel Primack and Uzy Smilansky
*Department of Physics of Complex Systems*
*The Weizmann Institute of Science, 76100 Rehovot, ISRAEL*

## Abstract

Non-generic contributions to the quantal level-density from parallel segments in billiards are investigated. These contributions are due to the existence of marginally stable families of periodic orbits, which are structurally unstable, in the sense that small perturbations, such as a slight tilt of one of the segments, destroy them completely. We investigate the effects of such perturbation on the corresponding quantum spectra, and demonstrate them for the stadium billiard.

# 1 Introduction

Hamiltonian systems which display strong forms of chaotic dynamics ($K$-systems) may tolerate families of periodic orbits which are marginally stable. They occupy a domain of measure zero in phase space and therefore do not affect the ergodicity of the system as a whole. These measure zero sets of periodic orbits are non-generic and also may be "structurally unstable" in the sense that an infinitesimal perturbation may destroy their periodicity completely. This sensitivity is in stark contrast with the structural stability of the chaotic component, which is robust against small perturbations. A notable example of such sets of trajectories are the "bouncing ball" (BB) orbits in the stadium billiard [1], which bounce perpendicularly between the parallel straight sections of the boundary. It is a continuous one parameter family of neutral periodic orbits, which disappear when the straight segments are slightly tilted.

Even though the non-generic components play insignificant rôle in the classical description, they might have rather noticeable effects on the spectral properties of the corresponding quantum analogues. Berry [2] studied their effect on the spectrum of the Sinai billiard. More recently they were brought again into the focus because of their prominent appearance in the spectrum of the stadium billiard [3, 4]: the Fourier transform of the spectral density (obtained numerically [4] and experimentally [3]) shows a sequence of structures which correspond to the multiple traversals of the bouncing ball orbits. This phenomenon was analyzed and explained in subsequent papers by Sieber et al [5] and by Alonso and Gaspard [6]. In view of the structural instability of these orbits, one could naively expect that the corresponding quantum features would also disappear as soon as the boundary is slightly deformed. However, quantum dynamics is a wave phenomenon, and as such it cannot resolve deformations which are smaller than the shortest wavelength in the relevant domain. Thus, it is more reasonable to assume that the quantum features which are due to non-generic and structurally unstable classical orbits will disappear gradually when the system is perturbed, and will leave their strongest mark on the higher parts of the quantum spectrum.

The theory in the next section, as well as the computational results of section 3, will explicitly demonstrate that the above expectation of gradual rather than abrupt transition actually holds. Also, a quantitative measure for the amount of deformation that is needed in order to observe the change in the quantum spectrum will be derived.

## 2 Theory

We shall start the discussion by deriving a general expression for the semiclassical contribution to the density of states function, produced by classical orbits that bounce from straight components of a planar billiard, exclusively. We shall apply this expression for billiards with parallel straight segments, such as e.g. the Bunimovich stadium [5], and the Sinai billiard [2]. The emphasis will be put on identifying the contributions that are generated by the classical periodic orbits. Then, the classical dynamics of orbits that bounce between two slightly tilted segments will be analyzed, and the results will serve to derive the resulting non-generic contribution to the density of states.

### 2.1 Level Density

The standard Gutzwiller's trace formula [7], which is frequently used to calculate the level density semiclassically for chaotic systems, does not account for the special effects of non-generic periodic orbits [5]. Therefore, we start our discussion from the expression for the more fundamental semiclassical propagator.

The contribution to the semiclassical propagator due to classical paths which reflect only from straight segments of the boundaries is given by:

$$K^s(\vec{q}\,', \vec{q}\,'', t) = \frac{m}{2\pi i \hbar t} \sum_{cl.paths} \exp\left[\frac{im}{2\hbar t} L^2(\vec{q}\,', \vec{q}\,'') - i\frac{\pi}{2}\nu\right] \quad (1)$$

where $m$ is the mass of the particle, $L$ is the length of the classical path from $\vec{q}\,'$ to $\vec{q}\,''$, and $\nu$ is a phase index, which is twice the number of collisions with the boundaries (for Dirichlet boundary conditions). It is assumed that between successive bounces, the motion is free.

The propagator can be Fourier-transformed to give the outgoing Green's function:

$$G^s(\vec{q}\,', \vec{q}\,'', E) \equiv \lim_{\epsilon \to 0^+} \frac{1}{i\hbar} \int_0^\infty K^s(\vec{q}\,', \vec{q}\,'', t) \exp\left\{\frac{i}{\hbar}(E + i\epsilon)\,t\right\} \mathrm{d}t \quad (2)$$

$$= \frac{m}{2i\hbar^2} \sum_{cl.paths} H_0^{(1)}(kL(\vec{q}\,',\vec{q}\,'')) \exp\{-i\frac{\pi}{2}\nu\} \qquad (3)$$

where $H_0^{(1)}$ is the Hankel function of order zero and first kind [8], and $k \equiv \sqrt{2mE}/\hbar$ is the wavenumber. Note, that for this special case, the integral in (2) could be calculated exactly, without invoking the stationary phase approximation.

The level density function is defined by:

$$d(E) \equiv -\frac{1}{\pi}\Im \int G(\vec{q},\vec{q},E) \mathrm{d}\vec{q} \qquad (4)$$

and we define for convenience the function $d(k)$, such that $d(k)\mathrm{d}k = d(E)\mathrm{d}E$. Inserting eq. (3) into (4) we get:

$$d^s(k) = \frac{\hbar^2 k}{m} d^s(E) = \frac{k}{2\pi} \int \sum_{cl.paths} J_0(kL(\vec{q},\vec{q})) \exp\{-i\frac{\pi}{2}\nu\} \mathrm{d}\vec{q} \qquad (5)$$

where the summation extends over all classical paths from $\vec{q}$ to $\vec{q}$ that reflect from straight segments only. Since $d(k)$ is a linear functional of the propagator $K$, one can identify $d^s(k)$ as the semiclassical contribution to the level density of those orbits that reflect from straight segments only.

## 2.2 Application to the Two-Parallel Lines Case

As a first application of (5), we shall study the case of two parallel lines of length $b$ and at a distance $a$, such as can be found in the stadium [1] (see fig. 1). The results of this calculation will serve as a reference point for the more general case to follow. In order to perform the calculation of $d^s(k)$, we should find for every point $\vec{q}$ inside the domain all the classical orbits that begin and end in $\vec{q}$. It is not difficult to see that these orbits are the ones that impinge vertically on the parallel lines. Actually, there is an infinite number of such orbits for every point $\vec{q}$, and we can divide them into two different classes:

1. Periodic or "even" orbits — these are the orbits whose final momenta are parallel to their initial momenta. They bounce an *even* number of times, and therefore have an overall phase of $+1$.

2. "Odd" orbits — these are the orbits whose final momenta are antiparallel to their initial momenta. They bounce an *odd* number of times, and their phase is -1.

The corresponding lengths of these orbits are:

$$L_e(n,x,y) = \left[(x_2 - x_1)^2 + (y_2 - y_1 + 2an)^2\right]^{1/2} , \; n \in \mathbf{Z} \qquad (6)$$

$$L_o(n,x,y) = \left[(x_2 - x_1)^2 + (y_2 + y_1 + 2an)^2\right]^{1/2} , \; n \in \mathbf{Z}. \qquad (7)$$

where we have explicitly used the Cartesian coordinates: $\vec{q} = (x,y)$. Inserting the above lengths into equation (5) we get after some manipulations:

$$d^s(k)_{parallel} = \frac{abk}{2\pi} \sum_{n=-\infty}^{+\infty} J_0(2akn) \; - \frac{b}{2\pi} \equiv d_e(k) + d_o(k) \qquad (8)$$

This result has been previously obtained by Berry [2], Sieber et al [5] and Alonso and Gaspard [6]. It differs from the generic contributions to $d(k)$ that come from unstable, isolated periodic orbits in two respects: firstly, it is composed of contributions from continuous families of neutral trajectories. Secondly, it includes contributions from orbits that do not close in *phase* space, but rather close in *configuration* (coordinate) space.

Analyzing eq. (8), we notice that the periodic (even) orbits give oscillatory terms in $k$. Actually, if $d_e(k)$ is Fourier-transformed [5], strong peaks are found for lengths $\pm 2ma$ ($m$ is a positive integer), corresponding to the classical periodic orbits. These peaks have also been observed in the microwave-cavity experiment [3] and in numerical simulations [4]. The $n = 0$ term in (8), which is the contribution of zero-length orbits, gives a smooth (non-oscillatory) contribution $abk/2\pi$. This is the leading (area) term in Weyl's asymptotic expression for the smooth level density in planar billiards [2, 9, 5] coming from the rectangle $a \times b$. The odd orbits yield a constant term, which is the contribution of the parallel lines to the first correction (perimeter) term in Weyl's formula.

There is a difference in powers of $k$ (and thus in powers of $\hbar$) between the contributions of the even and the odd orbits in (8). It is due to interference between the contributing amplitudes, which is *constructive* for the even orbits, and is *destructive* for the odd orbits. In the latter case there is an exact cancelation of the terms of order $\sqrt{k}$.

## 2.3 Classical Dynamics of the Deformed Billiard

The classical trajectories which bounce between the parallel segments are structurally unstable. An infinitesimally small tilt of the lines away from perfect parallelism, will cause the continuous family of periodic orbits to disappear. This, in turn, means, that had we used Gutzwiller's formula [7](or some other variant that relies upon periodic-orbit theory) to calculate $d(k)$, a discontinuous behavior (in some displacement parameter) would have shown up. In the following we shall show that the transition due to such a tilt is actually a continuous one.

We shall follow the behavior of the non-generic expression (8) when the straight segments are tilted at a small angle $\varphi$ (fig. 2). This justifies a first-order analysis, which makes the calculations tractable.

Once we have tilted the lines in the way described, the classical orbits which go from $\vec{q}$ to $\vec{q}$ will be deformed too: while in the parallel case the $x$-coordinate of the contributing orbits remains the same for all points of the orbit, it will be no longer the case in the deformed billiard, and the trajectories will stray from their initial $x$-coordinate.

Consider all the trajectories which start from an arbitrary point between the lines. It is easy to see, that if the trajectory initially aims at the narrowing part of the billiard, the scattering angle between the straight segment of the boundary and the trajectory will gain an angle $\varphi$ at every reflection from the lines. Therefore, the total scattering angle will eventually exceed $\pi/2$ and the trajectory will turn around towards the widening part of the billiard. This suggests, that there might be some initial launching angles, for which the trajectory will come back to its initial point, and a closed trajectory will be traced. If, however, the trajectory will initially go to the widening part, it will escape without ever coming back to its initial position.

We shall use elementary geometry to identify the angles for which the trajectories close. We shall first consider trajectories with initial momentum in the "upward" (negative $y$) direction. Since the scattering angle increases by $\varphi$ for every collision with the straight segments, it is not difficult to see, that a trajectory which starts with an angle of $m\varphi$ (where $m$ is a non-negative integer) relative to the negative $y$ axis, will scatter from the boundary at an angle $\pi/2$ after $m+1$ collisions. Hence, the trajectory will retrace itself, coming back to its initial position with final momentum that is opposite to the initial one. Fig. 3 shows such a trajectory, for $m=1$. Thus, the launching

angles:

$$\alpha_{o+}^n = n\varphi \, , \ n = 0, 1, 2, \ldots \qquad (9)$$

yield closed orbits. The subscript $o+$ stands for the fact, that these orbits substitute the odd orbits that start upwards in the parallel lines case. We note that the selected values of the launching angles are independent of the position $\vec{q}$, but for every point there is an upper limit for the launching angle (due to the finite extent of the straight segments). Also, no approximation had been used in obtaining this set.

In addition to the above family of closed orbits, we can find another (non-retracing) family that starts upwards (see fig. 3). Simple geometric and continuity arguments show, that between every two orbits with launching angles of $m\varphi$ and $(m+1)\varphi$, there is a unique closed orbit that closes from below. That is, its final momentum is approximately parallel ($\vec{p}_i \cdot \vec{p}_f > 0$) to the initial momentum. First–order (in the tilt angle) analysis gives the following formula for the launching angles:

$$\alpha_e^n = \varphi \frac{(na - y)}{a} \, , \ n = 1, 2, \ldots \qquad (10)$$

where the subscript $e$ stands for the fact, that these orbits substitute the even orbits in the parallel-lines case. Here the launching angle depends on the $y$ coordinate.

Until now we have considered only trajectories with initial momentum in the upward direction. If we include now trajectories with downward initial momentum, we get an additional family of odd-type trajectories:

$$\alpha_{o-}^n = n\varphi \, , \ n = 1, 2, \ldots \, . \qquad (11)$$

where the launching angle is now being measured relative to the positive $y$ direction. The absence of the $n = 0$ term in $\alpha_{o-}^n$ reflects the asymmetry of the coordinate system relative to the inclined lines, and does not have any special physical meaning. As regard the other family of orbits (eq. (10)), the uniqueness of the trajectories and forward-backward time symmetry implies that no new trajectories of this kind can be found. The same geometrical paths should be used, with the direction of propagation reversed.

First–order calculations in the tilt angle $\varphi$ give the following expressions for the lengths of the closed orbits:

$$L_{o+}^n = 2\{y + n[a + \varphi(b-x)]\} \ , \ n = 0, 1, 2, \ldots \tag{12}$$
$$L_{o-}^n = 2\{n[a + \varphi(b-x)] - y\} \ , \ n = 1, 2, 3, \ldots \tag{13}$$
$$L_e^n = 2n[a + \varphi(b-x)] \ , \ n = 0, 1, 2, \ldots \tag{14}$$

These expression are valid as long as the maximum excursion of the closed orbit in $x$ direction is small compared to $b$.

## 2.4 Calculation of the Non-Generic Contributions to the Level Density in the Deformed Billiard

We shall first study the contribution of the even family of orbits, and later will give some arguments, why the odd family can be ignored.

The calculation of $d_e(k)$ requires the application of eq. (5) to the family of orbits given in (14). However, special care should be given to the limits of the domain of integration in (5).

The relevant domain can be divided into 4 different parts, shown in fig. 2. Region 1 is the original $a \times b$ rectangle, and stands for the main contribution to the integral. Region 2 is formed due to the additional area in the deformed billiard relative to the original (parallel) one. Region 3 stands for the points in region 1, from which the classical trajectories eventually hit the chaotic scatterer. We should eliminate these points from the integration, since they no longer belong to the continuous family of closed orbits. Region 4 is composed of all the points that reside outside region 1, which are starting points of closed orbits which scatter only from the straight segments. The boundaries of regions 1 and 2 are independent of the number of collisions $n$ and the family of trajectories. The boundaries of regions 3 and 4 depend on both $n$ and the family of trajectories. They can be calculated (to first order in $\varphi$), and the result for the even family is:

$$\hat{x}_{even}^{(3)}(n, y) = \begin{cases} b + \varphi(\frac{y^2}{a} - y - \frac{an^2}{2}) & n \text{ even} \\ \\ b + \varphi(\frac{y^2}{a} - \frac{a(n^2+1)}{2}) & n \text{ odd} \end{cases} \tag{15}$$

$$\hat{x}_{even}^{(4)}(n,y) = \begin{cases} \varphi(yn - \frac{y^2}{a}) & y < \frac{an}{2n-1} \\ \varphi(an - (n-1)y - \frac{y^2}{a}) & y > \frac{an}{2n-1} \end{cases} \quad (16)$$

where $\hat{x}_{even}^{(3)}, \hat{x}_{even}^{(4)}$ are the boundary curves for regions 3, 4, respectively.

We consider first the contributions from regions 1 and 2. The contribution of the even orbits is calculated by inserting (14) into (5). Recalling the two families of even orbits, and the $+1$ phase generated by even number of bounces, it will be given by:

$$d_e(k) = 2 \sum_{n=1}^{N_e} \frac{k}{2\pi} \int_{x=0}^{b} dx \int_{y=0}^{a+\varphi(b-x)} dy \, J_0 \{2nk[a+\varphi(b-x)]\} \quad (17)$$

where $N_e$ stands for the number of valid terms to be included in this first-order analysis. Note that the domain of integration includes regions 1 and 2 only. The $y$ integration is trivial, and introduced a factor of $a+\varphi(b-x)$ (which is the local width) to the integrand. Changing variables to $t = a + \varphi(b-x)$, the integral can be performed analytically:

$$d_e(k) = \frac{b}{2\pi} \sum_{n=1}^{N_e} \frac{1}{n} \frac{(a+b\varphi)J_1[2nk(a+b\varphi)] - aJ_1(2nka)}{b\varphi}. \quad (18)$$

Each term in the sum is in the form of a "finite derivative" of the function $tJ_1(t)$ with respect to $b\varphi$. Taking the limit $\varphi \to 0$ and using the identity $tJ_0(t) = tJ_1'(t) + J_1(t)$, give us exactly the first term in (8), as expected.

The contributions from regions 3 and 4 are linear in $\varphi$. Since the main effect is dephasing between two leading-order terms (18) we can neglect the contributions of regions 3 and 4. This was validated by analytic and numerical calculations.

A better insight to the properties of $d_e(k)$ is gained by considering the length spectrum:

$$D_e(x, k_{max}) \equiv \frac{1}{k_{max}} \int_{k=0}^{k_{max}} \cos(kx) d_e(k) dk. \quad (19)$$

The domain of integration is taken to be finite (after [5]), in order to make it possible to relate the results to experiments. Inserting (18) into the above definition (19), and using the asymptotic expression for $J_1(x), x > 0$ [8] we get:

$$D_e(x, k_{max}) \simeq \sum_{n=1}^{N_e} \frac{1}{4\pi k_{max} n \varphi \sqrt{2\pi n}} \times$$
$$\left\{ \sqrt{a+b\varphi} \left[ \frac{S(k_{max}A^+) - C(k_{max}A^+)}{\sqrt{|A^+|}} - \frac{S(k_{max}A^-) + C(k_{max}A^-)}{\sqrt{|A^-|}} \right] \right.$$
$$\left. - \sqrt{a} \left[ \frac{S(k_{max}B^+) - C(k_{max}B^+)}{\sqrt{|B^+|}} - \frac{S(k_{max}B^-) + C(k_{max}B^-)}{\sqrt{|B^-|}} \right] \right\} \quad (20)$$

where we have defined:

$$A^\pm \equiv x \pm 2n(a+b\varphi)$$
$$B^\pm \equiv x \pm 2na$$

and:

$$S(x) \equiv \text{sign}(x) \int_0^{|x|} \frac{\sin(t)}{\sqrt{t}} \, dt$$
$$C(x) \equiv \int_0^{|x|} \frac{\cos(t)}{\sqrt{t}} \, dt.$$

Prominent peaks in (20) are expected for vanishing denominators: $x = \pm 2n(a+b\varphi)$, $x = \pm 2na$. Treating the $x > 0$ domain, we have the following picture. For very small $\varphi$, the two peaks are very large (of order $1/\varphi$) and very close (of order $\varphi$) to each other. Since $D_e$ in (20) has the structure of a "derivative", we expect to observe two sharp extrema with opposite signs near $x = 2na$, much like the case of differentiating a Gaussian or a Lorentzian. When $\varphi$ increases, the two peaks separate, their amplitude decrease, and we obtain destructive interference between the terms when:

$$2nk_{max}b\varphi \approx \pi. \quad (21)$$

This defines the critical tilt angle:

$$\varphi_{cr}^n = \frac{\pi}{2nk_{max}b} \quad (22)$$

for which we expect the structure of $D_e$ near $x = 2na$ to be appreciably lower and broader than for the $\varphi \to 0$ case.

A similar analysis had been applied to the odd family of orbits. The main result was, that $D_o(x, k_{max})$ is sharply concentrated near $x = 0$ for wide range of tilt angles, with no other prominent peaks. Since there are other contributions which are independent of $k$ (e.g., Weyl's term from the perimeter), it can be difficult and not practical to resolve the specific contribution of the odd family from all the other contributions.

## 3  Numerical Results and Discussion

In this section we shall demonstrate numerically the behavior of $D_e$ for a set of parameters that had been used by Gräf et al [3] to measure 1060 levels of the Bunimovich stadium, using superconducting microwave cavity. (The actual experiment was carried out in a parallel geometry, and we use the same parameters just for convenience.) The values of the parameters are: $a = 20$ cm, $b = 36$ cm, $k_{max} = 3.665$ cm$^{-1}$. Substituting these values into eq. (22) for the critical tilt angle we get:

$$\varphi_{cr}^n \approx \frac{0.012}{n}. \qquad (23)$$

In fig. 4 we plot $D_e(x, k_{max})$ (calculated from (20)) for three different values of $\varphi$, and for the term $n = 1$. For an extremely small angle, $\varphi = 10^{-6} \ll \varphi_{cr}^1$, we recover the results of Sieber et al. [5] for the parallel case, and the function has two prominent extrema with opposite signs near $x = 2a$ = 40 cm, as anticipated above. For $\varphi = 0.012 = \varphi_{cr}^1$ some lowering and broadening is observable. This trend is fully developed for $\varphi = 0.03$, where the peaks had been widen considerably, resulting in an amplitude which is 1/3 of the original one. Fig. 5 shows the heights of the positive peaks of $D_e(x, k_{max})$ as a function of $\varphi$ for $n = 1$ and $n = 2$ in (20) (the values are normalized according to the $\varphi = 0$ case). The decline of the maximal amplitude is evident, and the critical angle (23) (marked by cross) gives a good estimate for this decline to be already significant. We can also observe, that the decline for $n = 2$ is twice faster than for the $n = 1$ case. This is consistent with our prediction of the critical angle, eq. (23).

In summary, we obtained in this work a quantitative measure of the scale of deformations which are necessary to eliminate the effects of the bouncing ball orbits. This is of interest in the semiclassical study of billiards [5], where the non-generic families of orbits are a nuisance which masks and interferes with the contributions of the unstable periodic orbits. The effect can be demonstrated in microwave cavity experiments of the type reported in [3].

## Acknowledgement

We would like to thank Dr. Martin Sieber for many helpful discussions. This research was supported by grants from the U.S.-Israel bi-national science foundation (BSF) and the Israeli academy of science.

# Figures

**Fig. 1**: Geometry of the stadium billiard.

**Fig. 2**: Integration regions for the tilted stadium.

**Fig. 3**: Closed orbits in the tilted stadium. Right: odd orbit, left: even orbit.

**Fig. 4**: Length spectrum $D_e(x, k_{max})$ for $n = 1$ and tilt angle $\varphi = 10^{-6}$, 0.012 and 0.03 rad.

**Fig. 5**: Maximum of $D_e(x, k_{max})$ as a function of tilt angle $\varphi$ for $n = 1$ and $n = 2$. Values are normalized according to $\varphi = 0$ case and crosses mark $\varphi_{cr}^n$.

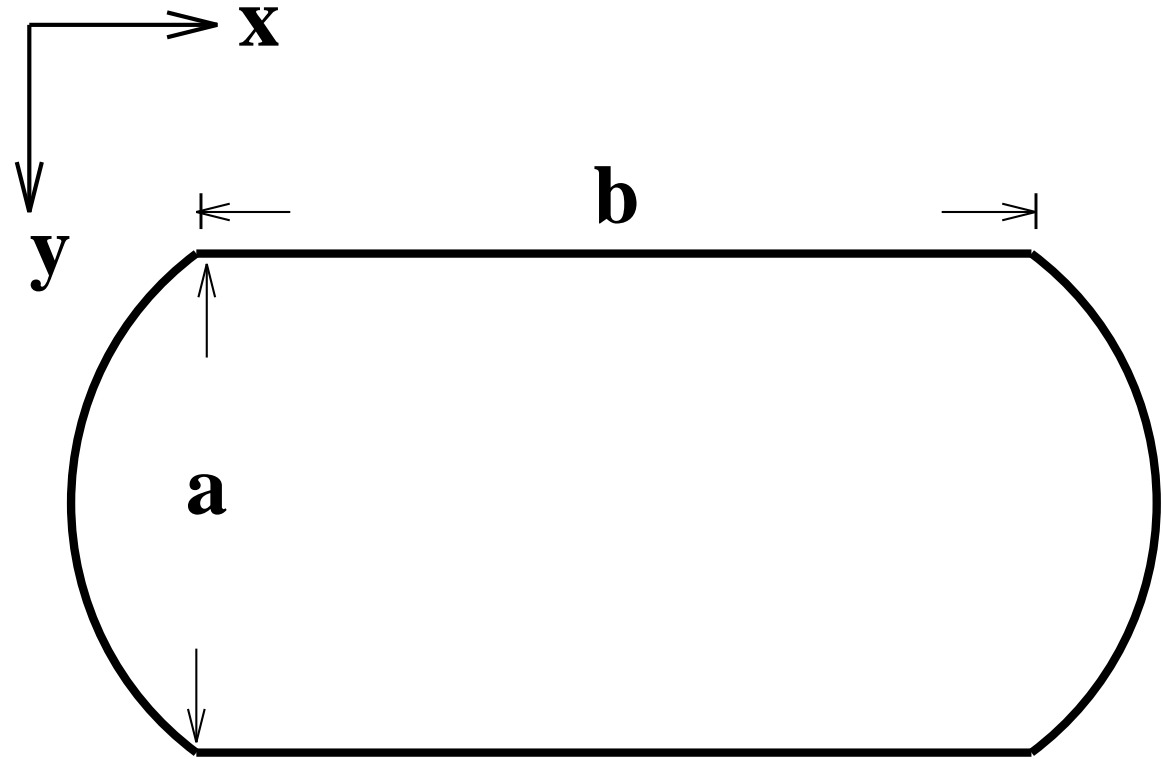

**Figure 1**

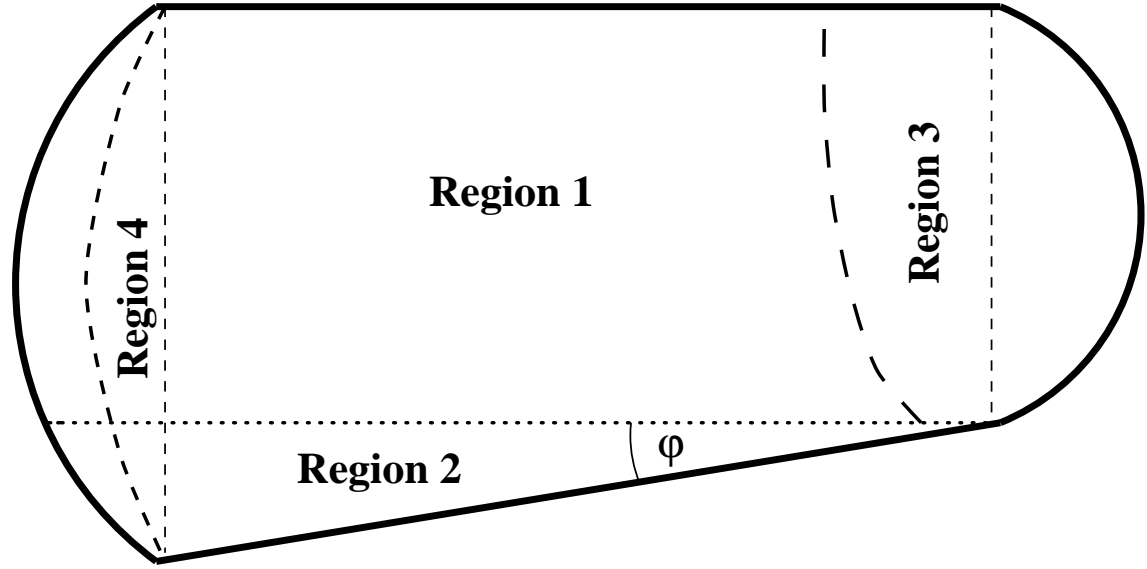

**Figure 2**

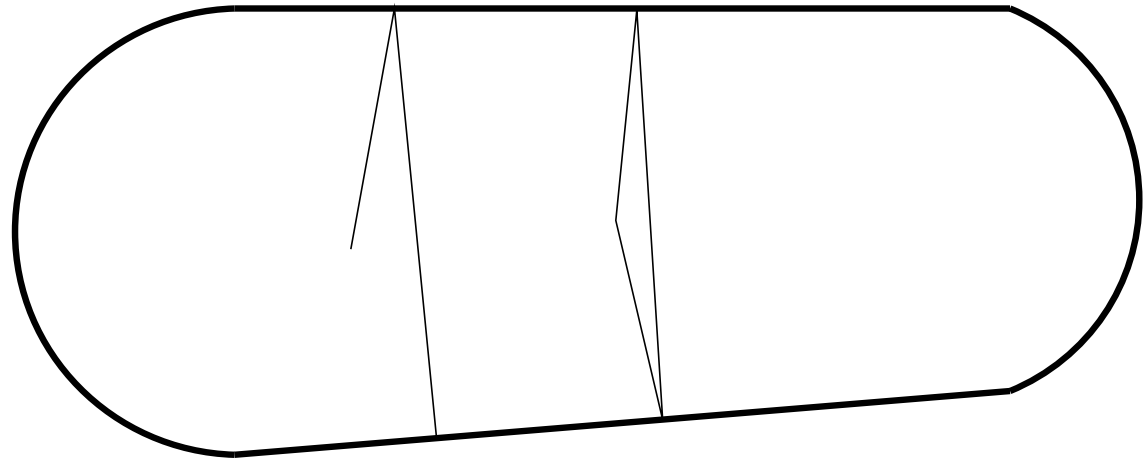

**Figure 3**

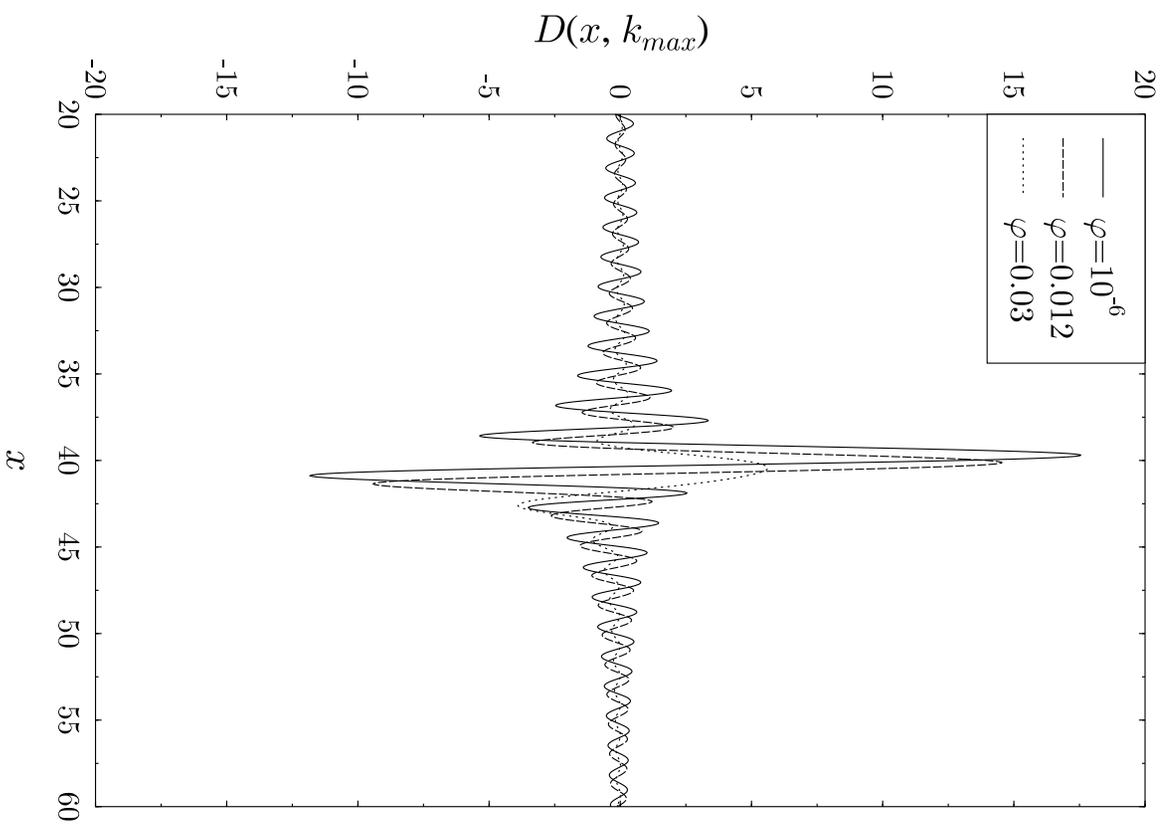

Fig. 4

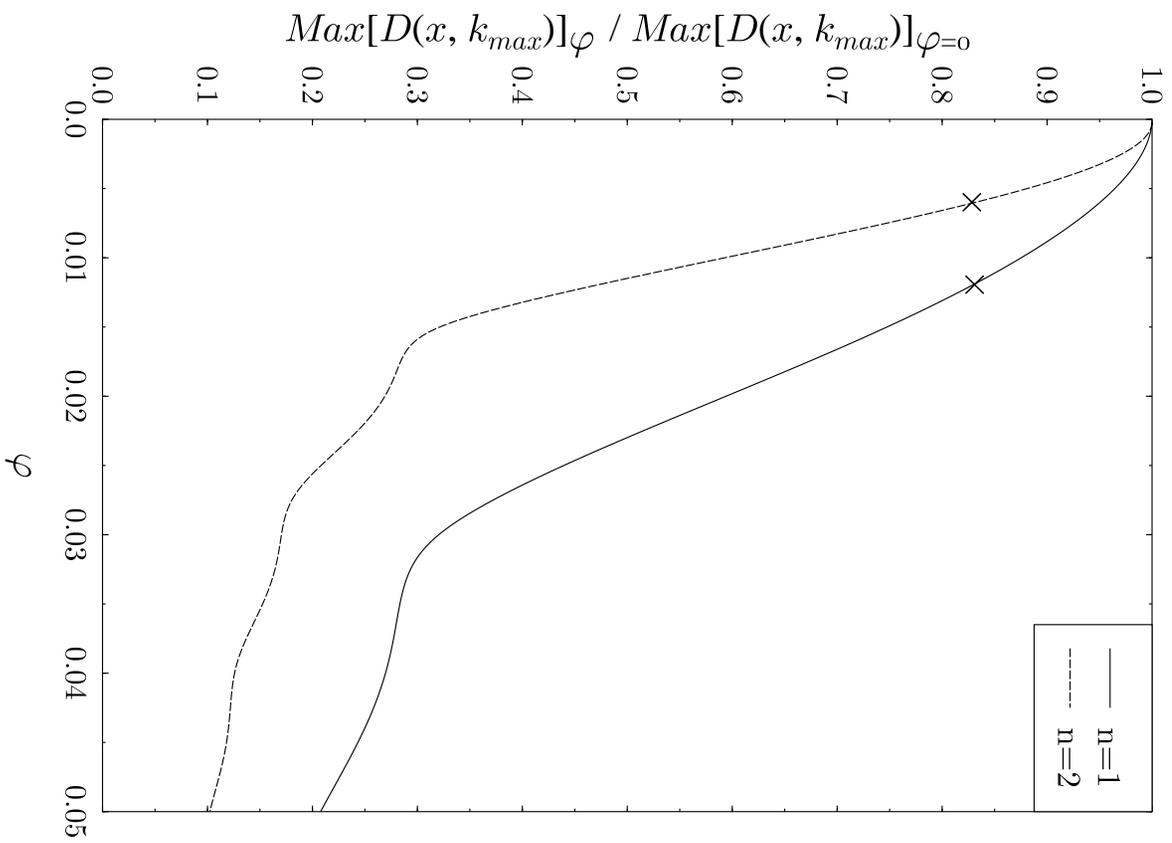

Fig. 5